\documentclass[journal,article,submit,pdftex,moreauthors]{Definitions/mdpi} 
\let\linenumbers\relax

\firstpage{1} 
\pubvolume{1}
\issuenum{1}
\articlenumber{0}
\pubyear{2025}
\copyrightyear{2024}
\datereceived{ } 
\daterevised{ } 
\dateaccepted{ } 
\datepublished{ } 
\hreflink{https://doi.org/} 

\Title{Quantum control of exciton motion in electric field}

\TitleCitation{Title}

\Author{Yingjia Li $^{1, 2, 3}$, Jorge Casanova $^{1, 3},$ Xi Chen $^{4}$ and  E. Ya. Sherman $^{1, 3}$}

\isAPAStyle{%
       \AuthorCitation{Lastname, F., Lastname, F., \& Lastname, F.}
         }{%
        \isChicagoStyle{%
        \AuthorCitation{Lastname, Firstname, Firstname Lastname, and Firstname Lastname.}
        }{
        \AuthorCitation{Lastname, F.; Lastname, F.; Lastname, F.}
        }
}

\address{%
$^{1}$ \quad Department of Physical Chemistry, University of the Basque Country UPV/EHU, Barrio Sarriena, s/n, 48940 Leioa, Spain \\
$^{2}$ \quad Institute for Quantum Science and Technology, Department of Physics, Shanghai University, 200444 Shanghai, China\\
$^{3}$ \quad EHU Quantum Center, University of the Basque Country UPV/EHU, Barrio Sarriena, s/n, 48940 Leioa, Spain\\
$^{4}$ \quad Instituto de Ciencia de Materiales de Madrid (CSIC), Cantoblanco, E-28049 Madrid, Spain}

\corres{Correspondence: evgeny.sherman@ehu.eus, laurynli36@gmail.com}

\abstract{We study quantum control of classical motion of a two-dimensional exciton by optimizing the time-dependent electric field of a stripe-like gate acting on the exciton and inducing its time-dependent quantum dipole moment. We propose a search method that significantly reduces computational requirements while efficiently identifying optimal control parameters.  By leveraging this method, one can precisely manipulate the exciton’s final position and velocity over a specified evolution time. These results can be applied for control of exciton fluxes and population, and for spatially resolved light emission in two-dimensional semiconducting structures.}


\begin{document}


\section{Introduction}
An exciton is a bound pair of an electron and a hole held together by Coulomb interaction \cite{knox1963theory}. In semiconductors, excitonic states are typically classified into two types. The first is the Frenkel exciton, characterized by a large binding energy and a short distance between the electron and hole, often localized on a single molecule. The second category is the Wannier-Mott exciton, which exhibits a smaller binding energy and a separation between the electron and hole significantly larger than the lattice constant. 

In two-dimensional (2D) materials, the bound states of electrons and holes form 2D excitons. In modern single-layer materials such as transition metal dichalcogenides, due to their ultrathin nature that results in reduced dielectric screening, excitons can exhibit relatively higher binding energies \cite{Wang2018} and remain bound even at room temperature. Moreover, 2D excitons are highly tunable under external electric and magnetic fields \cite{lozovik2002quasi}, making them a critical area of research in low-dimensional materials science. 

In semiconductors such as GaAs, excitons behave analogously to the proton-electron pairs in a 2D hydrogen atom  \cite{PhysRevA.43.1186}. The effective Bohr radius of an exciton is given by $a_\textrm{ex}=\epsilon m a_0/m^{*}= 11.86$ nm, where \( a_0 \) is the Bohr radius, \( \epsilon \) is the dielectric constant of GaAs,  $m$ denotes the electron mass, and \( m^{*} \) is the reduced mass of the electron-hole pair defined by $1/m^{*}=1/m^{*}_{e}+1/m^{*}_{h}.$ Here \( m_{e}^{*} \) and \( m_{h}^{*} \) are the effective masses of the electron and hole, respectively. The energy eigenvalues of exciton are expressed as
\begin{equation}
   E_
\textrm{ex} = -\frac{1}{(n - 1/2)^{2}} \frac{m^{*}}{m} \frac{1}{\epsilon^{2}} \frac{e^{2}}{2 a_0}, 
\end{equation}
where $e$ is the electron charge and \( n \) is the principal quantum number, an integer.

Excitons in semiconductors can be classified as direct or indirect based on whether the electron and hole reside in the same or different quantum well(s) (QW) or material layer(s) \cite{remeika2012two}. Indirect excitons, characterized by their extended lifetimes ranging from microseconds to milliseconds, facilitate more effective cooling and enable the formation of long-range coherent condensates under applied electric fields at low temperatures \cite{high2012spontaneous}. The electric field generated by split-gate electrodes plays a crucial role in controlling the transport of these indirect excitons \cite{dorow2018split} and enables the switching functionality of optoelectronic transistors \cite{high2007exciton}. By tuning the voltage applied to the electrodes, the resulting electric field distribution can be precisely modified, allowing for the manipulation of exciton energy and the creation of excitonic potential barriers essential for exciton collection and cooling \cite{kuznetsova2010control}. Additionally, the electric field of the gate facilitates the formation of interlayer excitons and regulates their transport within excitonic transistors, thereby significantly influencing the performance and efficiency of optoelectronic devices \cite{ ross2017interlayer}. Excitons trapped in a 2D lattice potential can be manipulated by varying the field direction or strength, enabling control over their transport and phase transitions \cite{remeika2012two}.

Furthermore, excitons offer innovative pathways for developing efficient and low-cost microscale and nanoscale optoelectronic devices. Excitonic solar cells are expected to become a key direction for high-efficiency solar cells in the future \cite{gregg2003excitonic}. Excitonic micro-LED fabricated on GaN nanowires have achieved record-high efficiencies for submicron scale LED \cite{pandey2023ultrahigh}, demonstrating the significant potential of excitons in advancing optoelectronic technology. These and others applications of excitons require a control of their position and velocities.  

In this paper we concentrate on direct excitons, where the electron and hole are located in the same high purity, high mobility quantum well. To achieve control over exciton dynamics, we employ a simple charged stripe-like gate. By modulating charge density on the  gate, we produce an electric field with spatially varying time-dependent distribution. 
This modulation of the electric field enables us to manipulate the quantum degree of freedom \cite{li2024quantum}, specifically the exciton dipole moment \cite{cohen1977quantum}, thereby controlling its classical motion in this field.

In more detail, when exciton is moving in an electric field, a relative displacement between the electron and hole occurs, inducing a dipole moment dependent on the strength and direction of the applied field, and, in a general case, the entire history of the previous exciton motion. As a result, in a nonuniform field a net force acting on the exciton appears and causes its motion. The control of classical motion can be achieved by modifying the gate charge altering the electric field and influencing the dynamics of the exciton via the time-dependent dipole moment. This control should satisfy the desired control tasks, e.g., in terms of the final position and velocity within a specified evolution time and given initial conditions. To achieve this control, we employ an optimal search method for the design of the gate charge producing the time-dependent quantum dipole moment of the exciton resulting in the realization of the control task for its classical motion. In our analysis we do not restrict ourselves to commonly used adiabatic approximation of slowly varying in time electric field. Thus, our approach is more robust and general.

The rest of the paper is organized as follows. In Sec. \ref{model}, we present the nonlinear model of the exciton quantum dynamics, consider self-consistently its classical motion, and outline the electric field of the gate. With this knowledge, we introduce in Sec. \ref{search method} the search problem, develop and detail an optimized search method, and present the search results for various dynamical regimes including nonadiabatic ones. Finally, we discuss possible relations to the experiment and formulate the conclusions of this paper in Sec. \ref{conclusions}.

\section{Model and equations of motion} 
\label{model}

\subsection{Evolution of Dipole Moment}
When an exciton, initially positioned at \(\bar{x}(0)\), that is the initial center-of-mass coordinate of the exciton, moves in an external electric field \( \mathcal{E}(\bar{x}(t)) \) oriented parallel to the \(x\)-axis, the system is described by the following Hamiltonian
\begin{equation}
H=H_0+H^{\prime}=H_0-{d} \cdot[\mathcal{E}(\bar x(t))-{\cal E}(\bar x(0))],
\label{Hamiltonian1}
\end{equation}
here, \(H_{0}\) denotes the Hamiltonian of the free two-level system \(\{|{g_0\rangle}, |{e_0\rangle}\}\), with energies \(E_{g}\) and \(E_{e}\), respectively, where we employ a qubit-like two-level model that can be used for a variety of different physical systems. It can be expressed as $H_0 = E_{g} |g_{0}\rangle\langle g_0| +E_{e} |e_{0}\rangle\langle e_0| =  \hbar \omega_0 \sigma_z/2$, where \(\sigma_z\) is the Pauli matrix, and the atomic transition frequency \(\omega_{0}\) is defined by \(\hbar \omega_0 = E_{e} - E_{g}\). The dipole moment is given by \({d} = -e (\hat{x} - \bar{x}(t))\) and is oriented parallel to the $x$-axis. Under the dipole approximation, the interaction Hamiltonian is $H^{\prime}=e(\hat{x}-\bar x(t))\left[\mathcal{E}(\bar x(t))-{\cal E}(\bar x(0))\right]$.

We expand the dipole moment operator in terms of the eigenstates of \(H_{0}\) as
\begin{equation}
{d} = {d_{ge}} |g_{0}\rangle\langle e_0| + {d_{eg}} |e_{0}\rangle\langle g_0|,
\end{equation}
where \({d_{ge}} \) is the matrix element of the dipole moment, and we assume \({d_{ge}} = {d_{eg}} ={\mu}\). At this point, the interaction Hamiltonian in Equation (\ref{Hamiltonian1}) is given by
\begin{equation}
H^{\prime} = -{d} \cdot \left[\mathcal{E}(\bar{x}(t)) - \mathcal{E}(\bar{x}(0))\right] = -{\mu} \sigma_x  \cdot \left[\mathcal{E}(\bar{x}(t)) - \mathcal{E}(\bar{x}(0))\right],
\end{equation}
where \(\sigma_{x}\) is the corresponding Pauli matrix. For this interacting system, we transform the Hamiltonian to the interaction representation, yielding \(H^{\prime}_I(t) = U_0^{\dagger}(t) H^{\prime} U_0(t)\), where $U_0(t) = \exp\left(-{i} H_0 t/{\hbar}\right) = \exp\left(-{i} \omega_0 \sigma_z t/{2}\right)$. All operators, including the Pauli, Hamiltonian, and dipole moment operators, are transformed into the interaction representation. For example, the dipole moment operator becomes $d(t) =d_{eg}(\sigma_x+i\sigma_y) e^{i \omega_0 t} /2+ \text{H.c.}$, where $\text{H.c.}$ denotes the Hermitian conjugate. Assuming the system’s wave function takes the form \( |\Psi(t)\rangle = a_e(t) |e_{0}\rangle + a_g(t) |g_{0}\rangle\), where $a_{e,g}(t)$ are complex coefficients, substituting this into the Schr\"{o}dinger equation \(i\hbar \, \partial_t |\Psi(t)\rangle = H_I |\Psi(t)\rangle\) yields a set of coupled partial differential equations
\begin{align}
  i\hbar \dot{a}_e(t) &= -{\mu} a_g(t)e^{i\omega_0 t} [\mathcal{E}(\bar{x}(t)) - \mathcal{E}(\bar{x}(0))],\label{ae} \\ 
  i\hbar \dot{a}_g(t) &=- {\mu} a_e(t)e^{-i\omega_0 t} [\mathcal{E}(\bar{x}(t)) - \mathcal{E}(\bar{x}(0))]\label{ag}.
\end{align}
Where $\dot{a}_g(t)$ and $\dot{a}_g(t)$ are derivatives with respect to \(t\).

At \(t = 0\), the total Hamiltonian in the interaction representation is
\begin{equation}
H = \begin{pmatrix} E_{g} & -\mu\mathcal{E}(\bar{x}(0))  \\ -\mu\mathcal{E}(\bar{x}(0)) & E_{e} \end{pmatrix}= \begin{pmatrix} E_{g} & V\\ V & E_{e} \end{pmatrix}.
\label{Hamiltonian3}
\end{equation}

The eigenvectors of the Hamiltonian given by Equation (\ref{Hamiltonian3}) are $ |\psi_+\rangle = \cos\left({\theta}/{2}\right) |g_{0}\rangle + \sin\left({\theta}/{2}\right) |e_{0}\rangle$ and $
  |\psi_-\rangle = \sin\left({\theta}/{2}\right) |g_{0}\rangle -\cos\left({\theta}/{2}\right) |e_{0}\rangle$, with the mixing angle $\theta \equiv \arctan[V/\Delta]$
and eigenvalues $E_{\pm} = E \pm \sqrt{(\hbar\omega_0/2)^{2} + V^{2}}$ where $
  E = (E_{g} + E_{e})/2.$ At this point, the expectation value of the dipole moment for the exciton in the ground state \( |\psi_-\rangle \) is given by 
\begin{equation}
    \langle d(0) \rangle =-e\langle{\psi_{-}|}\hat{x}-\bar x(t)|{\psi_{-}\rangle}=\mu \sin\left[\arctan\frac{2\mu\mathcal{E}(\bar{x}(0))}{\hbar\omega_0}\right]=
     \frac{2\mu^2\mathcal{E}(\bar{x}(0))}
    {\Delta(\bar{x}(0))},
    \label{induced_dipole1}
\end{equation}
where the gap $\Delta(\bar{x}(0))\equiv\sqrt{4\mu^{2}\mathcal{E}^{2}(\bar{x}(0)) + \hbar^{2}\omega_0^{2}}.$ Equation (\ref{induced_dipole1}) indicates that the induced dipole moment depends on the external electric field and does not vary linearly with the field. The expectation value of the time-dependent dipole moment in interaction representation is given by
\begin{equation}
    \langle d(t) \rangle =
-2 \operatorname{Re} \{d_{eg} e^{i \omega_0 t}a_e^{*}a_g\}. \label{exact_dipole}
\end{equation} 

Compared to \(H_{0}\), the term \(H^{\prime}\) is small and can be treated as a perturbation. The disturbance induced by the external electric field can be quantified using perturbation theory through energy corrections. For instance, when \(t = 0\), an exciton initially in the ground state at position \(a\) is subjected to the electric field \(\mathcal{E}(\bar{x}(0))\). If the external field is nonzero, it will generate an induced dipole moment. The Hamiltonian of the system in this case is given by $ H = H_0 + H^{\prime} = H_0 - d \cdot \mathcal{E}(\bar x).$

Using time-independent perturbation theory, the first-order correction eigenstates are $|{g}\rangle=|{g_0}\rangle-\mu\mathcal{E}(\bar x(0))/\hbar\omega_{0}|{e_0}\rangle$ and $|{e}\rangle={\mu \mathcal{E}(\bar x(0))}/{\hbar\omega_0}|{g_0}\rangle+|{e_0}\rangle.$ Thus, the expectation value of the dipole moment of the exciton in the ground state \(|{g}\rangle\) at \(t = 0\) is $\langle d(0)\rangle=-e\langle{g}|\hat{x}-\bar x(0)|{g}\rangle={2}\mu^2\mathcal{E}(\bar x(0))/{\hbar\omega_0}$.

When the external electric field is nonzero, or the initial velocity of the exciton is nonzero, the exciton will move within the field as time evolves. The Hamiltonian governing this motion is given by Equation (\ref{Hamiltonian1}). By analogy with standard time-dependent perturbation theory, we express \(|{\Psi(t)}\rangle\) as an expansion in terms of the previously obtained corrected eigenstates $|{\Psi(t)}\rangle=a_g(t)e^{-iE_{g}t/\hbar}|{g}\rangle+a_e(t)e^{-iE_{e}t/\hbar}|{e}\rangle$. Assuming the exciton initially occupies the lower state, the first-order approximation for the coefficients is given by 

\begin{equation}
a_g(t) = 1,\quad a_e(t) = -\frac{i}{\hbar}\int_{0}^{t}H_{eg}^{\prime}(t^{\prime})e^{i\omega_0t^{\prime}}dt^{\prime},
\end{equation}
where \(H_{eg}^{\prime}(t) \equiv \langle{e}| H^{\prime} |{g}\rangle\). The induced time-dependent dipole moment is then given by 

\begin{eqnarray}
&&\langle d(t)\rangle=-e\langle{\Psi(t)}|(\hat{x}-\bar x(t))|{\Psi(t)}\rangle
 \\&&=2\frac{\mu^2\mathcal{E}(\bar{x}(0))}{\hbar\omega_0}\cos(\omega_0 t)-2\frac{\mu^2}{\hbar}\cos(\omega_0 t)\int_{0}^{t}\sin(\omega_0 t^{\prime}){\cal E}(\bar x(t^{\prime}))dt^{\prime} \notag
 \\&& +2\frac{\mu^2}{\hbar}\sin(\omega_0 t)\int_{0}^{t}\cos(\omega_0 t^{\prime}){\cal E}(\bar x(t^{\prime}))dt^{\prime}. \notag
\label{dipole}
\end{eqnarray}

\begin{figure}[H]
    \centering
    \hspace*{-0.0cm} 
    \scalebox{0.65}[0.65]
    {\includegraphics{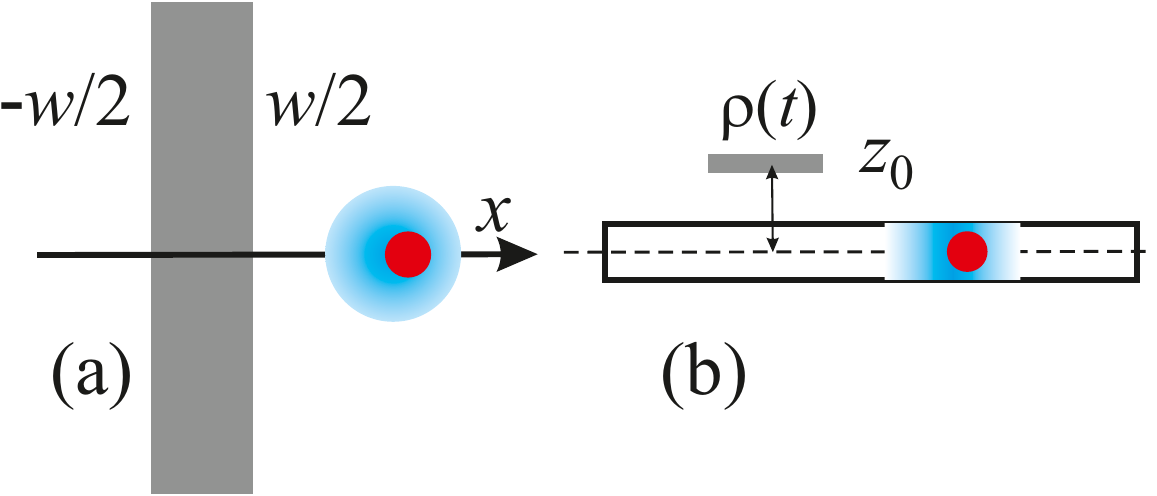}}
    \caption{(a) Top view of a stripe-like gate (gray stripe) with charge density \(\rho(t)\) and polarized exciton in the field of the stripe; (b) Side view of the stripe-like gate, showing the exciton located at a distance \(z_{0}\) below the gate. }
    \label{stripe-like gate}
\end{figure}

At \(t = 0\), the time-dependent dipole moment in Equation (\ref{dipole}) is $\langle d(0)\rangle={2}\mu^2\mathcal{E}(\bar x(0))/{\hbar\omega_0}$, which agrees with the result obtained from the time-independent perturbation theory. 

\subsection{Classical Equation of Motion}

In our approach the position variable $\bar{x}(t)$ denotes the center-of-mass coordinate of the exciton, while the external electric field $\mathcal{E}(\bar{x}(t))$ acquires a time dependence in the exciton's eigenframe even if the gate charge is time-independent. As an exciton is a composite quantum particle with a finite spatial extent, we adopt the Ehrenfest approach, which treats its center-of-mass motion classically and, in our case, controllable by acting on the quantum degree of freedom, that is, the dipole moment.

The energy of a dipole in an external field is given by \(-d(t) \cdot \mathcal{E}(\bar{x}(t))\). Spatial variations in the potential energy (i.e., the gradient of the potential) induce the force acting on exciton resulting in its acceleration, expressed as \(M \ddot{\bar{x}}(t) = -\nabla U(\bar{x}, t) = (d \cdot \nabla) \mathcal{E}\), where \(M = m_{e}^{*} + m_{h}^{*}\) is the mass of the exciton. By incorporating the dipole moment expression in Equation (\ref{exact_dipole}), the system's dynamical behavior can be described by 
\begin{equation}
M\ddot{\bar{x}}(t)=\frac{\partial \mathcal{E}(\bar{x}(t))}{\partial \bar{x}(t)}\langle d(t) \rangle.
\label{newton} 
\end{equation}
Henceforth, unless specified, we omit the $t$-dependence for brevity, e.g. $a_{g,e}(t) \equiv a_{g,e}$. By combining Newton's equation of motion, Equation (\ref{newton}), with the dipole moment expression, Equation (\ref{exact_dipole}), and coupling them with Equations (\ref{ae}) and (\ref{ag}), we can numerically solve the system under the initial conditions $a_e(0)=-\cos\left({\theta}/{2}\right)$ and $a_g(0)=\sin\left({\theta}/{2}\right)$. This approach allows us to determine the exciton's trajectory in the external electric field and the induced time-dependent dipole moment.

Under the adiabatic approximation of slow time evolution of the electric field, the dipole moment follows the electric field $\mathcal{E}(\bar{x}(t))$ as described by Equation (\ref{induced_dipole1}). The corresponding classical equation of motion in adiabatic limit is given by 

\begin{equation}
   M\ddot{\bar{x}}(t)=\frac{\partial \mathcal{E}(\bar{x}(t))}{\partial \bar{x}(t)}\frac{2\mu^2\mathcal{E}(\bar{x}(t))}{\sqrt{4\mu^{2}\mathcal{E}(\bar{x}(t))^{2} + \hbar^{2}\omega_0^{2}}}.
\label{adiabatic limit} 
\end{equation}
The slow motion of the exciton requires the electric field to vary sufficiently slowly, ensuring that the probability of non-adiabatic transitions between instantaneous eigenstates is very small. We present this condition in the form $d\Delta(\bar{x}(t))/dt\ll \Delta^{2}(\bar{x}(t)),$ where $\Delta(\bar{x}(t))$ is the instantaneous gap. Here $d\Delta(\bar{x}(t))/dt=2\Delta^{-1}(\bar{x}(t))\mu^{2}d\mathcal{E}^{2}(\bar{x}(t))/dt$ with 
$d\mathcal{E}^{2}(\bar{x}(t))/dt=\partial\mathcal{E}^{2}(\bar{x}(t))/\partial\,t+\bar{v}(t)\partial\mathcal{E}^{2}(\bar{x}(t))/\partial x.$ The latter expression implies that the adiabaticity 
condition depends both on the change of the electric field with time as determined by the charge of the gate and by exciton velocity.

Now we introduce the units. The electron charge \(e = -1\), 
and the reduced Planck constant $\hbar=1.$ The units for frequency, mass and dipole moment are taken as \(\omega_0= M= \mu = 1\). Consequently, the unit of length is defined as \(\langle g_0|\hat{x}|e_{0}\rangle\) for the exciton with $\bar{x}=0.$ For an exciton in GaAs, the characteristic length is approximately $10\,\text{nm}$, and the corresponding transition frequency is on the order of $10^{12}\, \text{s}^{-1}$.

\subsection{Electric Field of the Gate}
As shown in Figure \ref{stripe-like gate} (a), we model the gate by a uniformly charged stripe infinitely extended in the $y$-direction $ -\infty < y < \infty $ and having a width $w$ along the $x$-direction. The total electric field at point ${x}$ can be decomposed into contributions from the central region (\(|y| \leq L\)) and distant regions (\(|y| > L\)) as: 
\begin{equation}
\mathcal{E}(x) = \rho\int_{-\infty}^{\infty} { \cos\theta}\frac{dy}{r^2} = \rho\int_{-L}^{L} { \cos\theta}\frac{dy}{r^2} + 2\rho\int_{L}^{\infty} { \cos\theta}\frac{dy}{r^2},    
\end{equation}
where $r$ is the distance from a gate element $dy$ to the exciton, and $\cos\theta$ accounts for the component of the electric field along the $x$-axis. The contribution of the distant regions is simplified as $ 2\rho\int_{L}^{\infty} { \cos\theta}/{r^2} dy\sim\rho\int_{L}^{\infty}{x w}/{y^3}dy= {x\rho w}/{L^2}$. For modern systems with $L$ exceeding 1 $\mu$m and typical values of $x$ considered here being of the order of 100 nm, this contribution is negligibly small and can be ignored. The resulting electric field in Gaussian units at the point $x$ is
\begin{equation}
{\mathcal{E}}(x)=\frac{\rho(t)}{\epsilon}\ln{\frac{({w}/{2}+x)^{2}+z_{0}^{2}}{({w}/{2}-x)^{2}+z_{0}^{2}}}.
\label{electric_field}
\end{equation}
Below we define $\bar\rho(t)\equiv\rho(t)/\epsilon$ and $z_{0}$ denotes the distance along the $z-$axis between the exciton and the gate.

\begin{figure}[t]
    \centering
  \scalebox{0.42}[0.42]{\includegraphics{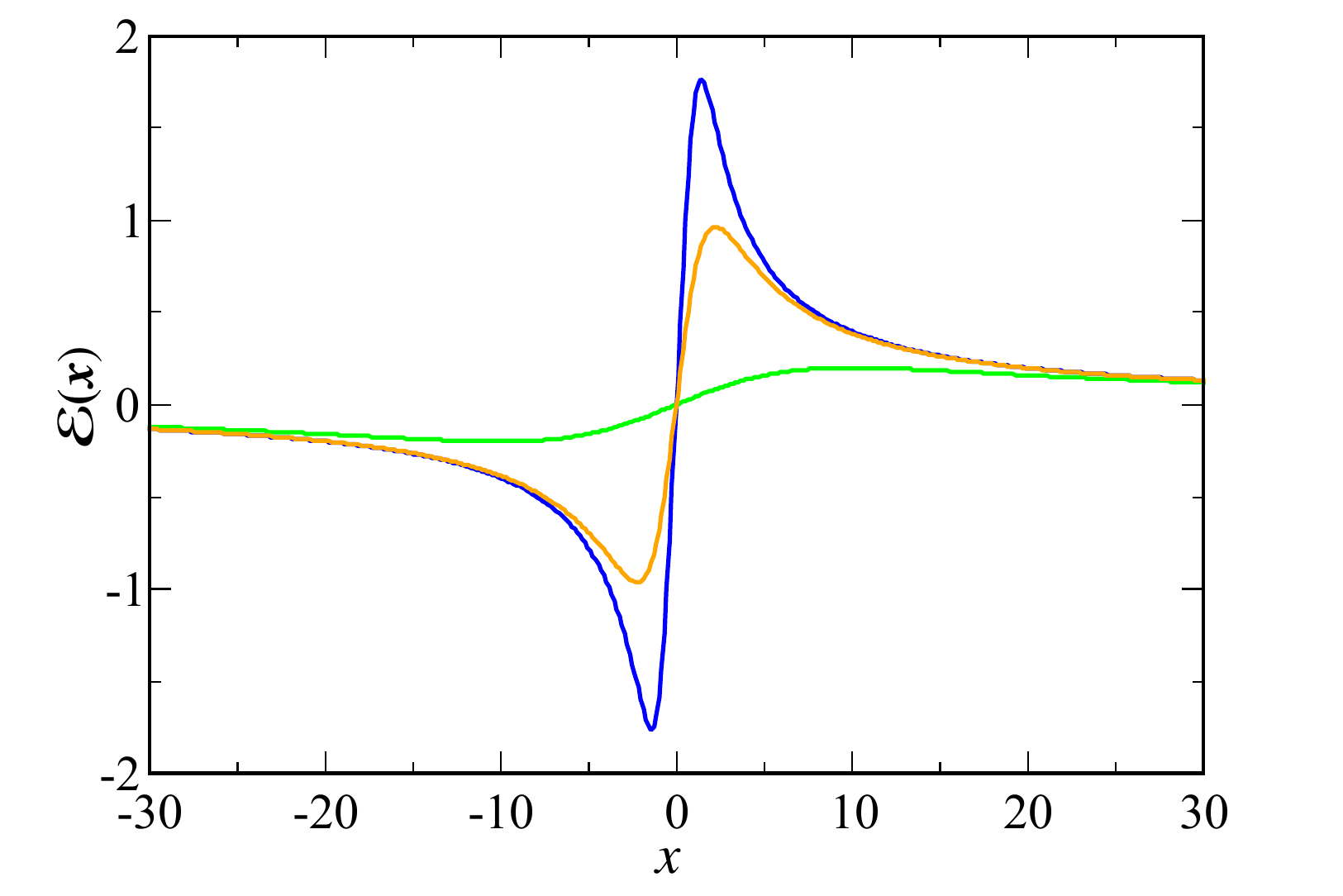}}
    \caption{Variation of electric field with $x$ for different $z_{0}$. Blue, orange and green lines represent the distances $z_{0}=1$, $z_{0}=2$ and $z_{0}=10,$ respectively. The gate width $w=2.$ In the SI units, an electric field value of $\mathcal{E}(x) = 1$ corresponds to about $10^3\,\text{V/cm}$, and position and gate width values of order one corresponds to about 10 nm.}
    \label{electric field}
\end{figure}

Figure \ref{electric field} illustrates the electric field produced by the gate when the charge density is constant, \(\bar\rho(t) = 1\), at various positions $x.$ Near the center of the gate, the electric fields from either side almost cancel each other resulting at $|x|\ll w$ and $|x|\ll z_{0}$ in $\mathcal{E}(x)\approx 2\bar\rho wx/(z_{0}^{2}+w^{2}/4).$  At $|x|\gg w$ and $|x|\gg z_{0}$ one obtains $\mathcal{E}(x)\approx 2\bar\rho w/x.$ Figure \ref{electric field} represents the $x-$dependent electric field demonstrating that its decrease at a large distance is $z_{0}-$ independent.

\section{Control of classical motion}
\label{search method}
\subsection{Search Problem}
\label{search problem}

The motion of an exciton in an external electric field depends on the initial velocity \(\bar{v}(0)\) and position \(\bar{x}(0)\) and responds dynamically to variations in the electric field. 
Since we are interested in the quantum control of classical motion, we formulate the conditions for the final position and velocity without specifying the final quantum state as:
\begin{equation}
x_{1,f}<\bar{x}(t_{f})<x_{2,f};\qquad v_{1,f}<\bar{v}(t_{f})<v_{2,f}, 
\label{conditions}
\end{equation}
with the initial conditions $\bar{x}(0)$ and $\bar{v}(0).$ Thus, our objective is to control the exciton's final position and velocity by modifying the external electric field. 

For the sake of generality, we mention the relation of this problem to other aspects of quantum control and quantum mechanics. This problem is considerably different from the set of problems studied in the shortcuts to adiabaticity \cite{Muga2019} since we consider the effect of the quantum dynamics on the classical motion, not considered in the shortcuts to adiabaticity approaches. However, this problem can be seen as an inverse classical version of the time-of-arrival problem in the fundamentals of quantum mechanics \cite{beau2024time}.  In addition, it can be related to the problem of quantum backflow \cite{Palmero2013,Barbier2021} where the interaction of the quantum motion of the electron and hole can lead to modification of the classical flux of excitons.

\subsection{Search Method}

Here, we produce a time-dependent electric field by modulating the charge density
\(\bar\rho(t)\). To control \(\bar\rho(t)\), we adopt the first kind Chebyshev polynomial Ansatz $\bar\rho(t) = \sum_{j=0}^{n}c_{j}T_{j}(t)$, subject to the boundary conditions \(\bar\rho(0) = \bar\rho(t_{f}) = 0\). 
The first kind Chebyshev polynomials defined as \( T_j(x) = \cos(j \arccos x) \) are closely related to trigonometric functions, making it possible to approximate complex functions with relatively simple polynomials. For instance, \( T_0(x) = 1 \), \( T_{1}(x) = x \), \( T_{2}(x) = 2x^{2} - 1 \), \( T_3(x) = 4x^3 - 3x \), and \( T_4(x) = 8x^4 - 8x^{2} + 1 \), illustrate how higher-degree polynomials can be constructed recursively. The coefficients \(c_j\) are determined by applying the Ansatz in conjunction with the specified boundary conditions. For \(n = 4\), the pulse-shaped charge density distribution can be adjusted using three free parameters \(c_{0}, c_{1},\) and \(c_{2}\). Imposing $\bar\rho(0)=0$ and $\bar\rho(t_{f})=0$ allows $c_3$ and $c_4$ to be expressed in terms of $c_{0}$, $c_1$, and $c_2$, yielding
\begin{equation}
\label{density}
\bar\rho(s)= c_{0} T_0(s) + c_{1} T_{1}(s) + c_{2} T_{2}(s) - (c_{1} + 2c_{2}) T_3(s) - (c_{0} - c_{2}) T_4(s), 
\end{equation}
with $s=t/t_{f}.$ By adjusting various parameter combinations, such as \(c_{0}\),  \(c_{1}\) and \(c_{2}\) in Eq. (\ref{density}), we can guide the exciton's motion through the resulting electric field.
\begin{figure}[t]
  \centering
  \scalebox{0.6}[0.6]{\includegraphics{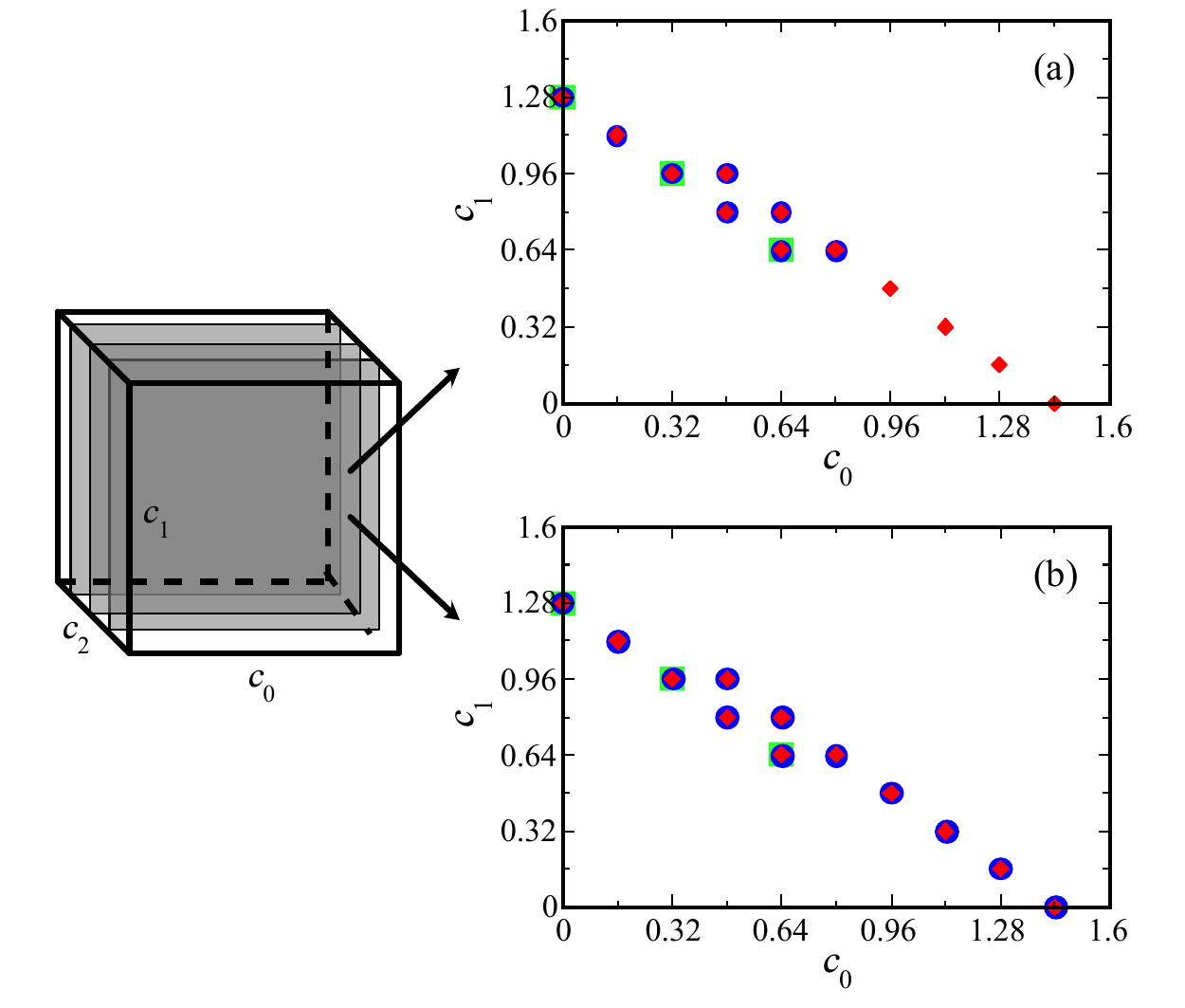}}
  \caption{
Slice of the $(c_{0},c_{1})$ plane at \(c_{2} = 1.6\). (a) shows the slice obtained using the original search method. (b) presents the results after applying the modified search method with an expanded search range. The green squares represent valid points identified by direct calculation with \(\delta_c = 0.32\), the red rhombus indicates valid points found by direct calculation with \(\delta_c = 0.16\), and the blue circles denote valid points identified by the search method with \(\delta_c = 0.16\). Parameters: \(\bar{v}(0) = 0\), \(\bar{x}(0) = 1\), \(z_{0} = 5\), \(w = 2\), \(t_{f} = 10\), \(x_{1,f} = 8\), \(x_{2,f} = 8.5\), \(v_{1,f} = 1.1\), and \(v_{2,f} = 1.2\). In the SI units, a velocity of order one in our model corresponds to \(10^4\, \text{m/s}\), and a time unit corresponds to \(1\, \text{ps}\).}
  \label{slice_demo}
\end{figure}

To address the search problem defined above, we focus on identifying combinations of three parameters \(c_{0}, c_{1},\) and \(c_{2}\) under a certain initial condition. The combinations we are expected to identify should satisfy two final conditions. Firstly, the exciton’s final position falls within the interval \([x_{1,f}, x_{2,f}]\). Secondly, the the final velocity falls within \([v_{1,f}, v_{2,f}]\). The variables satisfy these conditions are defined as "valid points" while the "invalid points" are the combinations that failed to satisfy the conditions.

This search problem, involving three parameters, can be considered as a three-dimensional (\(3D\)) problem, thus can be mapped as a cube in three dimensional space, as illustrated in Figure \ref{slice_demo}. The three edges of the cube correspond to the parameters \(c_{0}\), \(c_1\), and \(c_2\), each with a length of \(10.24\), indicating that the the range of the parameters is $[0,10.24]$.  Each point within the cube signifies a unique combination of \(c_{0}\), \(c_1\), and \(c_2\). By discretizing these parameters using the specified step size \(\delta_c\), we can systematically evaluate whether each point is valid or not within the cube.

To identify the valid points, the most straightforward approach is to directly calculating all points. Although this full-calculation strategy ensures comprehensive exploration, it becomes computationally inefficient for highly complex systems. The inefficiency turns severe when the degrees of freedom is large, or the parameter discretization is high. The computational time may increase exponentially, and a higher portion of the calculating resources might be wasted on irrelevant or inefficient regions. In addition to the direct calculation method, other methods could increase efficiency while losing some precision of identification. 

Therefore, the efficient method we expect to develop has to increase the computational efficiency and keep precision. To achieve this, we outline the following steps. 

\begin{itemize}
\item	Step 1: We begin with selecting a relatively larger initial step size $\delta_c$ and employing direct calculation. For each point within the cube, the corresponding charge density varies accordingly. Using the Runge-Kutta method, we numerically solve the coupled Equations (\ref{ae}), (\ref{ag}), (\ref{exact_dipole}), and (\ref{newton}). This process yields the exciton's final position \(\bar{x}(t_{f})\) and final velocity \(\bar{v}(t_{f})\) at \(t = t_{f}\). Each point satisfied the conditions in Equation (\ref{conditions}) is labeled as valid, others are labeled as invalid.
\end{itemize}

Here Step 1 only requires relatively less time and resources to get a rough outline of the distribution of valid points. However, a more granular identification of valid points is essential for further analysis. Therefore, the following steps attempt to narrow the search range by only searching for potential points near the valid points under a smaller \(\delta_c\).

\begin{itemize}
\item Step 2: Each valid point obtained from Step 1 is now treated as an initial point to construct a local neighborhood. Toward an initial point \(\mathcal{P}_0 (c_{0}, c_{1}, c_{2})\), we define its neighborhood \(\mathcal{N}_1\) by independently adjusting each coordinate $c_i$ $(i = 0, 1, 2)$ with a reduced step size \(\delta_c/2\): 
\begin{equation} \label{N1}
\mathcal{N}_1(\mathcal{P}_0, \delta_c/2) = \{c_{0} + \Delta_0\,{\delta_c}/{2},\; c_{1} + \Delta_1\,{\delta_c}/{2},\; c_{2} + \Delta_2\,{\delta_c}/{2}\},     
\end{equation}
where the multipliers $\Delta_0$, $\Delta_1$, and $\Delta_2$ are chosen from the set $\{-1, 0, 1\}$, which corresponds to decreasing the coordinate by $\delta_c/2$, leaving it unchanged, or increasing it by $\delta_c/2$, respectively. This approach generates all possible combinations of such adjustments simultaneously for $c_{0}$, $c_1$, and $c_2$.
This yields \(3^D = 27\) total combinations (\(D=3\)), one being \(\mathcal{P}_{0}\) itself and the remaining \(3^D - 1\) points surrounding it;

\item Step 3: Applying the same numerical method used in Step 1, we evaluate the points in each neighborhood to determine which point remains valid under the reduced step size $\delta_c/2$. By focusing on neighborhoods around previously identified valid points, we concentrate computational effort on the regions with potential to find the valid points; 

\item Step 4: Valid points from Step 3 serve as updated initial points. The step size is further reduced to \(\delta_c/4\), and the same process is repeated along all \(3^D - 1\) directions. By iteratively narrowing the step size and concentrating on neighborhoods around valid points, we achieve a more precise and efficient identification of additional valid parameter combinations.
\end{itemize}
 
The described search method significantly reduces the calculation of the points and then improves computational efficiency. Intuitively, we define efficiency improvement as the ratio of the number of computational points reduced by the search method to the number of computational points of direct calculation.

\begin{table}[H] 
\caption{ Comparison of parameter evaluations for the adiabatic case $t_f=10$, valid point counts, and efficiency improvement for the \textit{original search method} versus \textit{direct calculation} at varying step sizes \( \delta_c \). For each \( \delta_c \), the table lists the number of parameter combinations evaluated and the number of valid points found (in parentheses) by both methods, along with the corresponding percentage of efficiency improvement. The results demonstrate that the search method identifies a high proportion of valid points while dramatically reducing the number of evaluations required compared to exhaustive direct calculation. Parameters are the same as Figure \ref{slice_demo}.}
\begin{tabularx}{\textwidth}{CCCC}
\toprule
\textbf{$\delta_c$}	& \textbf{Original Search Method}	& \textbf{Directly Calculating} & \textbf{Efficiency Improvement}\\
\midrule
0.32 & 32768 (18) & 32768 (18)  & 0\% \\
0.16 & 327 (105) & $0.26*10^{6}$ (131)& 87.27\% \\
0.08 &  1591 (799)  & $>0.2*10^7 (989)$  & $\approx 98.27\%$ \\
\bottomrule
\end{tabularx}
\label{1}
\end{table}

Table \ref{1} presents a comparison of the computational efficiency between the search method and direct calculation, utilizing the final conditions \(x_{1,f} = 8\), \(x_{2,f} = 8.5\), \(v_{1,f} = 1.1\), and \(v_{2,f} = 1.2\) for excitons in an external electric field. On the first round of searching, as mentioned in Step 3, the search method calculated 131 parameter combinations and identified 105 valid points, achieving $78.95\%$ of the valid points found by direct calculation. These 105 points served as the initial points for a subsequent search round, where 1,591 combinations were evaluated, resulting in 799 valid points. In contrast, direct calculation requires evaluating over two million points to identify all 989 valid points. Thus, the search method successfully identified $80.79\%$ of the total valid points while improving computational efficiency by approximately $98.27\%$.

Although the search method significantly enhances computational efficiency, it may overlook valid points, particularly those located further away from the initial valid points. This limitation arises from the neighborhood-based approach, which focuses on evaluating points within a fixed local region around each initial point. As illustrated in Figure \ref{slice_demo} (a), green squares represent valid points identified through direct calculation with \(\delta_c = 0.32\), which served as the initial points for subsequent searches. However, the search method evaluates only 26 directions around each initial point, with a reduced step size of \(\delta_c = 0.16\). In the slice of the $(c_{0},c_{1})$ plane, the blue dots denote the valid points discovered by the first round of searching, while the red rhombus represents valid points found by direct calculation with $\delta_c=0.16$. The missing points are located in the lower right region of the slice, indicating areas the search method fails to reach when initialized with the green squares. Consequently, these omissions occur in subsequent search iterations, limiting the method's overall effectiveness.

To address this limitation, it becomes necessary to expand the search range. To enhance the coverage, the neighborhood \(\mathcal{N}_2\) is extended in all directions, allowing the search to encompass \((D+2)^3 - 1\) combinations with a reduced step size of \(\delta_c/2\), similar to Equation \eqref{N1} as: 
\begin{equation}
\mathcal{N}_2(\mathcal{P}_0, \delta_c/2) = \{c_{0} + \Delta_0 \delta_c/2, c_{1} + \Delta_1 \delta_c/2, c_{2} + \Delta_2 \delta_c/2\}.
\end{equation}
In this definition, the multipliers \(\Delta_0\), \(\Delta_1\), and \(\Delta_2\) are chosen independently from the set \(\{\pm 1, 0, \pm 2\}\), indicating that each coordinate \(c_i\) is adjusted by integer multiples of \(\delta_c/2\). This expanded neighborhood increases the discretized search points in all directions within the cube, thereby mitigating the risk of missing valid points located further away from the initial points. Consequently, Steps 3 and 4 of the search method are adjusted to accommodate the expanded scope.

In Step 3, the points within the extended neighborhood \(\mathcal{N}_2(\mathcal{P}_0, \delta_c/2)\) are evaluated using the same numerical methods outlined in Step 1. This extension allows the identification of valid points located further from the initial points obtained in Step 1, thereby improving the overall coverage of the search space. In Step 4, the newly identified valid points from Step 3 are used as updated initial points. The step size is then reduced to \(\delta_c/4\), and the search process is repeated within the further refined neighborhood \(\mathcal{N}_2(\mathcal{P}_0, \delta_c/4)\), ensuring a more thorough exploration of the parameter space.
\subsection{Search Results}
We maintain the initial and final conditions of excitons in an external electric field and apply the modified search method to identify valid points anew. As shown in Table \ref{2}, after the first round of searching (with a step size of $\delta_c/2$), the modified search method identified 129 valid points. In the second round of searching (with a step size of $\delta_c/4$), a total of 984 valid points were identified. Compared to the original search method, this refined neighborhood approach increased the proportion of valid points found in the two search rounds by $19.52\%$ and $18.70\%$, achieving $98.47\%$ and $99.49\%$ of the total valid points, respectively. 

Figure \ref{slice_demo} (b) illustrates the $(c_{0},c_{1})$ plane at \(c_{2} = 1.6\) obtained using the modified search method. While the original method, with its limited search range, omits many valid points, especially those further from the initial green squares, the expanded search successfully identifies these previously missed points, including those in the lower right region of the same slice.

\begin{table}[H] 
\caption{Efficiency comparison between the \textit{modified search method} with an expanded search range and \textit{direct calculation} for $t_f=10$ at different step sizes $\delta_c$. Each row corresponds to a specific $\delta_c$ and shows the number of parameter combinations evaluated and valid points found (in parentheses) using the modified search method versus direct calculation and the percentage of efficiency improvement. The data illustrate how the modified search method maintains a high proportion of valid points while significantly reducing computational effort. Parameters are the same as Figure \ref{slice_demo}.}
\begin{tabularx}{\textwidth}{CCCC}
\toprule
\textbf{$\delta_c$}	& \textbf{Modified Search Method}	& \textbf{Directly Calculating} & \textbf{Efficiency Improvement}\\
\midrule
0.32 & 32768 (18) & 32768 (18) & 0\% \\
       0.16 & 903 (129) & $0.26*10^{6}$ (131) & 87.05\% \\
      0.08 &  4221 (984) & $>0.2*10^7 (989)$& $\approx 98.11\%$ \\
\bottomrule
\end{tabularx}
\label{2}
\end{table}

This expanded neighborhood approach significantly enhances the effectiveness of the search method by addressing its primary limitation, ensuring a higher proportion of valid points are identified while maintaining computational efficiency. This modified search method effectively solves such optimization problems and can significantly conserve computational resources.

 \begin{figure}[H]
  \centering
  \scalebox{0.65}[0.65]{\includegraphics{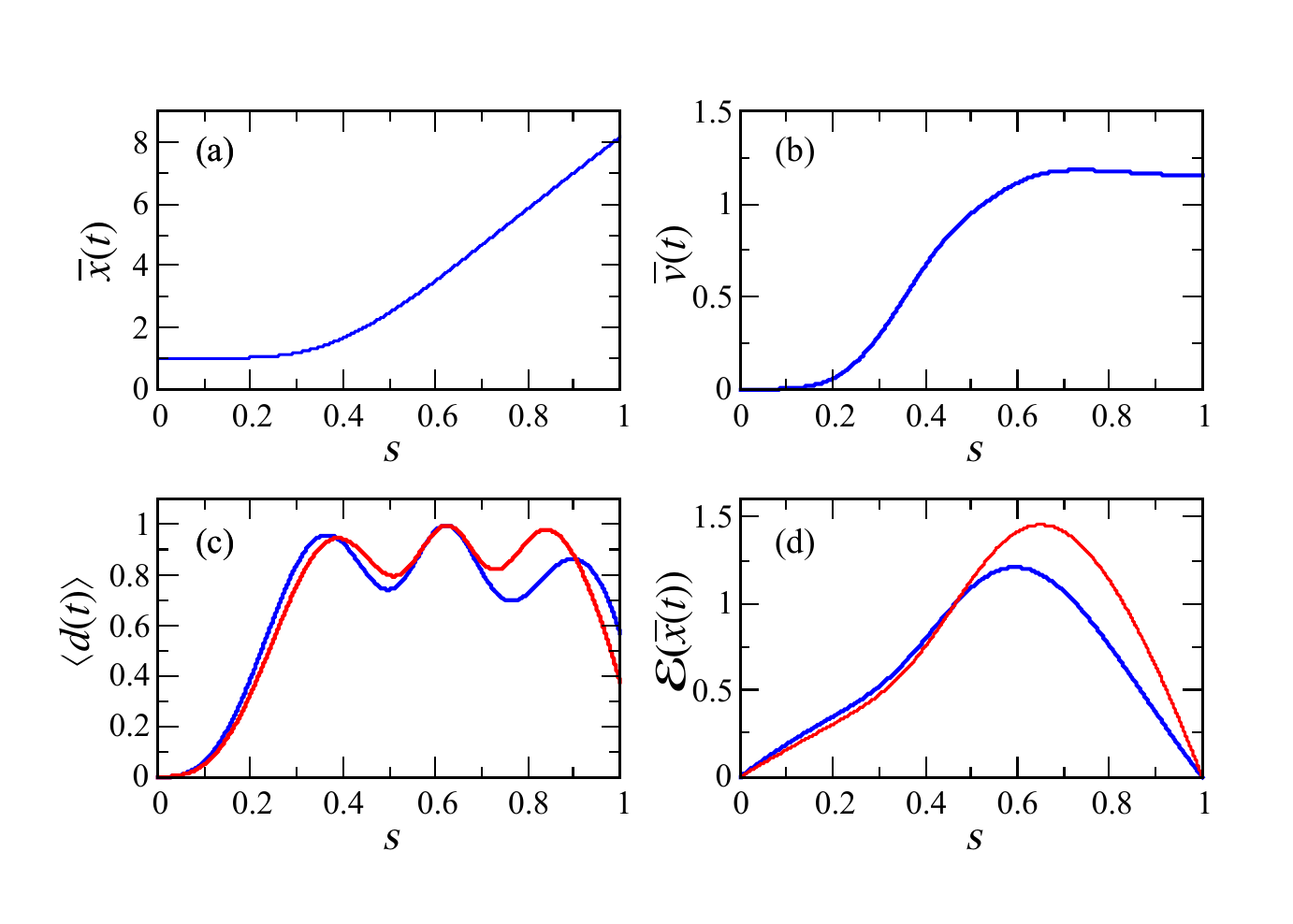}}
  \caption{Time evolution of (a) exciton position, (b) velocity, (c) dipole moment, and (d) electric field for two distinct charge density parameter sets. The red solid lines correspond to \(c_{0} = 0.32\), \(c_{1} = 1.92\), and \(c_{2} = 0.48\), while the blue solid lines represent \(c_{0} = 0.32\), \(c_{1} = 0.96\), and \(c_{2} = 1.6\). Other parameters are the same as Figure \ref{slice_demo}. Panels (a) and (b) show only the blue line since it almost coincides with the red one.}
  \label{evolution1}
\end{figure}

Figure \ref{evolution1} shows how the exciton's position, velocity, dipole moment, and external electric field evolve over time under a stripe-like gate with modulated charge density. The solid blue and red lines correspond to two different sets of charge density parameters: \(c_{0} = 0.32\), \(c_{1} = 0.96\), \(c_{2} = 1.6\) and \(c_{0} = 0.32\), \(c_{1} = 1.92\), \(c_{2} = 0.48\), respectively. Figures \ref{evolution1} (a) and (b) present the evolution of $\bar{x}(t)$ and $\bar{v}(t)$, both showing smooth increases. Figure \ref{evolution1} (c) plots the dipole moment $\langle d(t)\rangle$, which grows with the electric field initially and around the expected value of 1 when $\omega_0 t_{f} \gg 1$ in the adiabatic regime. This value corresponds to the ground state \(|\psi_-\rangle = (|g_{0}\rangle - |e_{0}\rangle)/\sqrt{2}\), where the diagonal energy terms are relatively small. The dipole moment oscillates around this value due to quantum effects. Figure \ref{evolution1} (d) illustrates the external electric field \(\mathcal{E}(\bar{x}(t))\), demonstrating the robustness of the optimized electric field in precisely controlling the exciton's motion. 

\begin{figure}[t]
  \centering
  \scalebox{0.65}[0.65]{\includegraphics{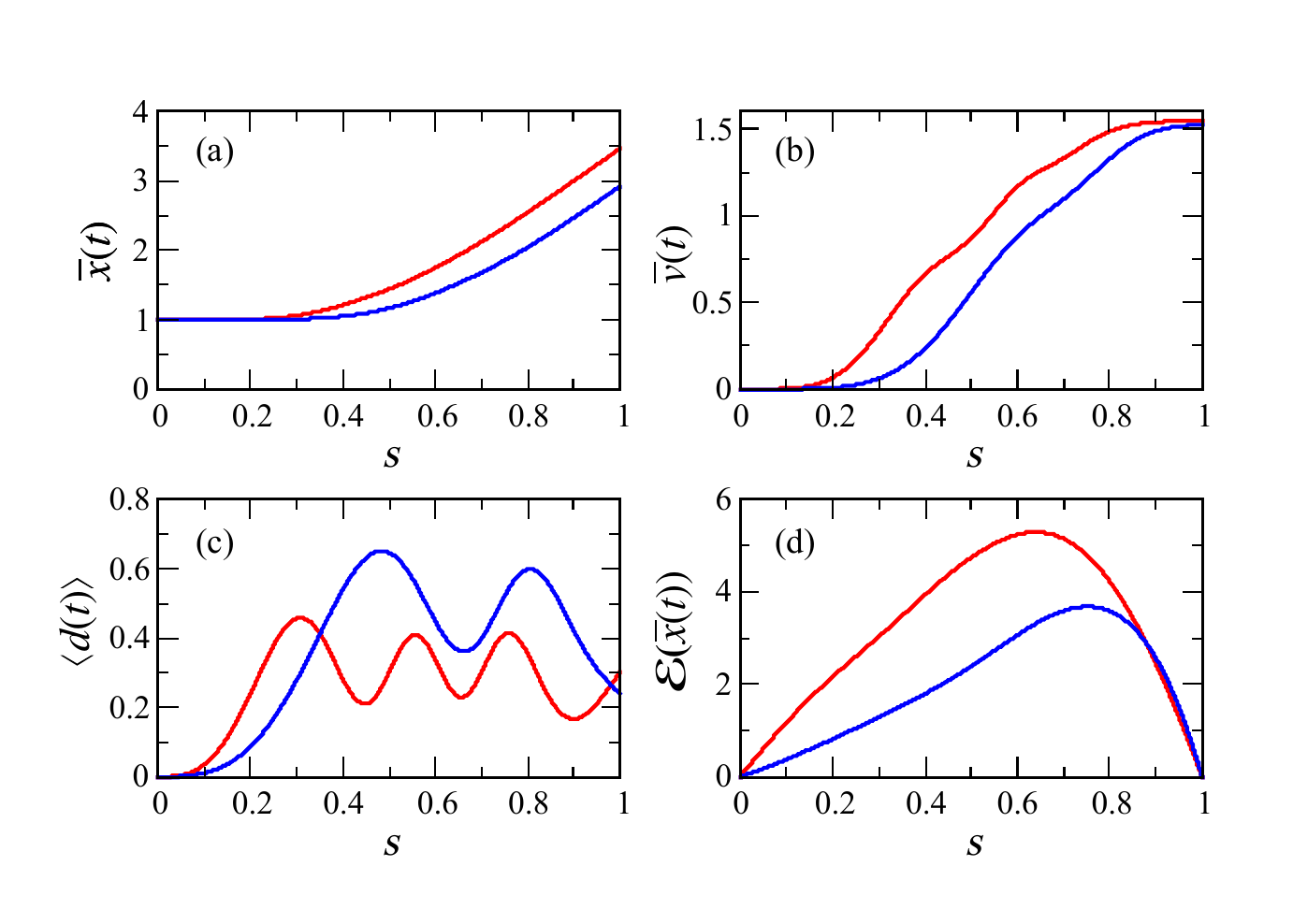}}
  \caption{Time evolution of (a) exciton position, (b) velocity, (c) dipole moment, and (d) electric field for two distinct charge density parameter sets. The red solid lines correspond to \(c_{0} = 1.6\), \(c_{1} = 8\), and \(c_{2} = 8.64\), while the blue solid lines represent \(c_{0} = 6.4\), \(c_{1} = 0.8\), and \(c_{2} = 3.2\). Parameters: $\bar{v}(0)=0$, $\bar{x}(0)=1$, $z_{0}=5$, $w=2$, $t_{f}=3$, $x_{1,f}=3.5$, 
$x_{2,f}=4$, $v_{1,f}= 1.5$ and $v_{2,f}= 1.6$.}
  \label{evolution3}
\end{figure}

Next, we consider a relatively short \(t_{f} = 3\), where the adiabatic picture is not applicable. When \(t_{f} = 3\), for two selected sets of parameters \(c_{0}, c_{1}, c_{2}\), the time evolution of the exciton's position, velocity, dipole moment, and the external electric field is presented in Figure \ref{evolution3}. The velocity \(v(t)\) in Figure \ref{evolution3} (b) oscillates more strongly, indicating less smooth variation over shorter times. In Figure \ref{evolution3} (c), the dipole moment $\langle d(t) \rangle$ initially follows the electric field. However, due to the rapid variation of the field, which violates the adiabatic condition, \(\langle d(t) \rangle\) exhibits significant oscillations and no longer strictly follows the field evolution, resulting in a noticeable deviation from the adiabatic evolution. Consequently, at the final time \(t = t_{f}\), the dipole moment does not decrease in tandem with the diminishing electric field, starkly contrasting with the adiabatic evolution scenario.

\begin{table}[H] 
\caption{Comparison of the \textit{modified search method} and \textit{direct calculation} at various step sizes $\delta_c$ under nonadiabatic conditions at $t_f=3$. For each $\delta_c$, the table provides the number of parameter combinations evaluated, the count of valid points obtained (in parentheses) by both methods and the percentage efficiency improvement achieved by the modified search method. The table highlights the modified method's effectiveness in rapidly identifying all valid points while substantially cutting down on required computations. Other parameters are the same as Figure \ref{evolution3}.}
\begin{tabularx}{\textwidth}{CCCC}
\toprule
\textbf{$\delta_c$}	& \textbf{Modified Search Method}	& \textbf{Directly Calculating} & \textbf{Efficiency Improvement}\\
\midrule
0.32 & 32768 (111) & 32768 (111) & 0\% \\
       0.16 & 3471 (977) & $0.26*10^{6}$ (985)  &  86.06\% \\
       0.08 & 18618 (8178 )  & $>0.2*10^7 (8203) $  & $\approx  97.26\%$ \\
\bottomrule
\end{tabularx}
\label{3}
\end{table}

\begin{figure}[H]
  \centering
  \scalebox{0.65}[0.65]{\includegraphics{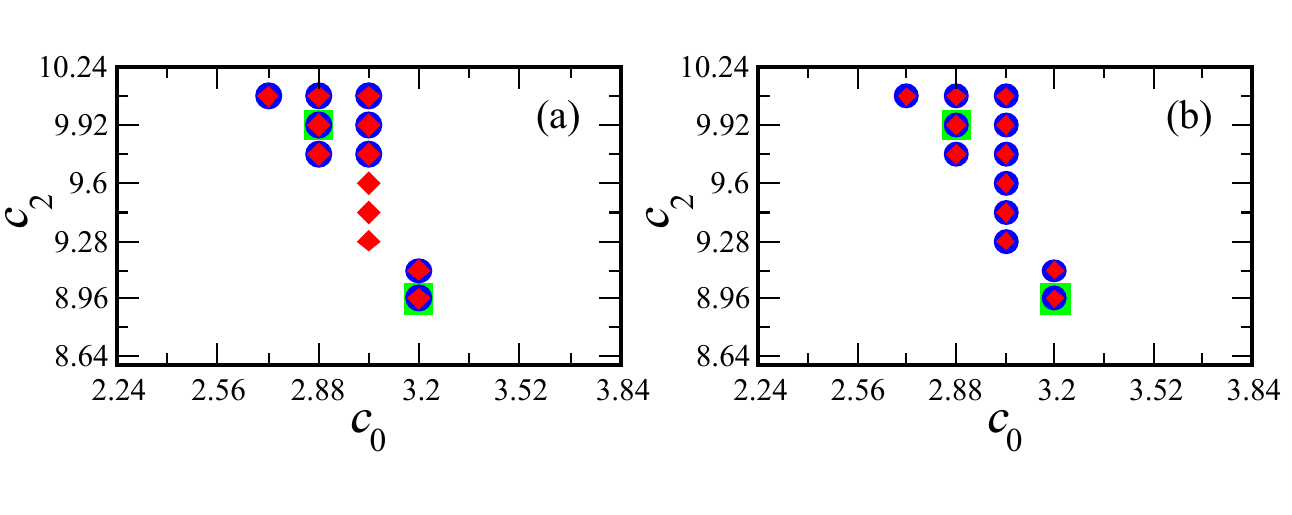}}
  \caption{Slice of the $(c_{0},c_{2})$ plane at \(c_{1} = 6.4\). (a) shows the slice obtained using the original search method. (b) presents the results after applying the modified search method with an expanded search range. The green squares represent valid points identified by direct calculation with \(\delta_c = 0.32\), the red rhombus indicates valid points found by direct calculation with \(\delta_c = 0.16\), and the blue circles denote valid points identified by the search method with \(\delta_c = 0.16\). Other parameters are the same as Figure \ref{evolution3}.}
  \label{slice_comparition_tf3}
\end{figure}

Figure \ref{slice_comparition_tf3} compares the results of the original and modified search methods on the same $(c_{0},c_{2})$ plane at \(c_{1} = 6.4\). The original search method, shown in Figure \ref{slice_comparition_tf3} (a), is constrained by a limited search range and consequently identifies fewer valid points, particularly missing those further from the initial points. In contrast, the modified search method, as depicted in Figure \ref{slice_comparition_tf3} (b), significantly enhances coverage by expanding the search range, thereby successfully locating additional valid points, including those initially missed. As summarized in Table \ref{3}, the search method identifies 977 valid points in the first search round, corresponding to \(99.19\%\) of the total 985 valid points. In the second search round, the method successfully identifies 8,178 valid points, accounting for \(99.70\%\) of the total 8,203 valid points. This demonstrates the improved efficiency and accuracy of the modified search method in identifying valid parameter combinations within a reduced timeframe.  

\section{Conclusions and outlook}
\label{conclusions}

We proposed an efficient algorithm of quantum control of the classical motion of excitons in a high-quality semiconductor quantum well, based on the design of time- and position-dependent electric fields produced by a stripe-like gate. The exciton is considered in a two-level \textit{qubit-like} model permitting to study main features of its motion while the time dependence of the field is not restricted to the adiabatic approximation. The electric field  of the gate generates a time- and position-dependent dipole moment of the exciton resulting in a net force acting on it and producing its classical acceleration. The ways to improve the efficiency of the search algorithm are presented and analyzed. 

The proposed approach can be used for the control of exciton energy transfer processes and optics of semiconducting quantum wells by various local static and time-dependent electric fields. In general, it can be applied to the analysis of dynamics of qubit-like two-level systems such as spins of single electrons and excitons in the presence of spin-orbit coupling and system magnetization \cite{Sadreev2013,Ungar2019}. The proposed control protocols can be extended to multilevel exciton models similar to qutrits and their generalizations. In addition, this approach can be seen as a supplemental one in operational optimization of quantum dots-based devices \cite{ji2024surface} such as $\text{AgBiS}_{2}$ solar cells where excitonic effects play a significant role.

\vspace{6pt} 



\authorcontributions{Conceptualization, E.S. and X.C.; methodology, E.S.; software, Y.L.; formal analysis, Y.L., E.S. and J.C.; investigation, Y.L., X.C. and J.C.; writing---original draft preparation, Y.L.; writing---review and editing, E.S., X.C., and J.C.; visualization, Y.L.; supervision, E.S.; funding acquisition, X.C. and J.C.  All authors have read and agreed to the published version of the manuscript.}

\funding{Supported by Grant PID2021-126273NB-I00 of MCIN/AEI/10.13039/501100011033 and ERDF “A way of making Europe” and by the Basque Government through Grant No. IT1470-22 and IKUR STAQC project. J. C. acknowledges the Ram\'{o}n y Cajal (RYC2018-025197-I) research fellowship. Authors acknowledge the Quench project that has received funding from the European Union's Horizon Europe -- The EU Research and Innovation Programme under grant agreement No 101135742, the Spanish Government via the Nanoscale NMR and complex systems project PID2021-126694NB-C21.}

\conflictsofinterest{The authors declare no conflicts of interest.}

\abbreviations{Abbreviations}{
The following abbreviations are used in this manuscript:\\

\noindent 
\begin{tabular}{@{}ll}
2D & two-dimensional \\
QW &  quantum well(s)
\end{tabular}
}

\appendixtitles{no} 

\reftitle{References}

\bibliography{dipole.bib}

\end{document}